\documentclass[fleqn,usenatbib]{mnras}
\usepackage{booktabs}
\usepackage{newtxtext,newtxmath}

\usepackage[T1]{fontenc}

\DeclareRobustCommand{\VAN}[3]{#2}
\let\VANthebibliography\thebibliography
\def\thebibliography{\DeclareRobustCommand{\VAN}[3]{##3}\VANthebibliography}

\usepackage{graphicx}	
\usepackage{amsmath}	

\title[Super-Jeans fragmentation]{Super-Jeans fragmentation in massive star-forming regions revealed by triangulation analysis}

\author[Guang-Xing Li et al.]{
Guang-Xing Li,$^{1}$\thanks{gxli@ynu.edu.cn, ligx.ngc7293@gmail.com}
Mengke Zhao,$^{2,1}$
Xing Lu$^{3}$
\\
$^{1}$South-Western Institute For Astronomy Research, Yunnan University, Kunming 650600, China\\
$^{2}$School of Astronomy and Space Science, Nanjing University, 163 Xianlin Avenue, Nanjing 210023, People’s Republic of China\\
$^{3}$South-Western Institute For Astronomy Research, Yunnan University, Kunming 650600, China\\
}
\date{Accepted XXX. Received YYY; in original form ZZZ}

\pubyear{\the\year{}}

\begin{document}
\label{firstpage}
\pagerange{\pageref{firstpage}--\pageref{lastpage}}
\maketitle

\begin{abstract}
	{Understanding the fragmentation of the gas cloud and the formation of massive stars remains one of the most challenging questions of modern astrophysical research. Either the gas fragments in a Jeans-like fashion, after which the fragments grow through accretion, or the fragmentation length is larger than the Jeans length from the start. Despite significant observational efforts, a consensus has not been reached. The key is to infer the initial density distribution upon which gravitational fragmentation occurs.
Since cores are the products of the fragmentation process, the distances between adjacent cores serve as a scale indicator. Based on this observation, we propose a Delaunay triangulation-based approach to infer the density structure before the fragmentation and establish the link between density distribution and gas fragmentation length. We find that at low density, the fragmentation is Jeans-like, and at high densities, the core separations are larger than the prediction of the Jeans fragmentation. This super-Jeans fragmentation is a key step toward the formation of massive stars.}
\end{abstract}

\begin{keywords}
    galaxies: star formation -- ISM: clouds -- Interstellar Medium: Nebulae --    methods: data analysis 
\end{keywords}



\section{Introduction }\label{sec1} 
\noindent 
Stars form from the collapse of gas clouds, which are complex, multi-scale systems. 
Understanding how such a system evolves remains one of the most difficult challenges in astrophysical research.
 Massive stars have masses larger than the Jeans mass of clouds. The Jeans mass \citep{1902RSPTA.199....1J} is the characteristic mass of gravitational instability, and to explain the formation of massive stars, one needs to explain how massive stars can acquire masses much larger than the Jeans mass. 
There are two possibilities: The first \citep{2001MNRAS.323..785B} assumes that massive stars start with cores whose masses are comparable to the Jeans mass. Since the Jean mass is small, these cores must accrete significant mass before becoming stars. 
Another possibility is to form cores whose masses are much larger than the Jeans mass from the start \citep{2019MNRAS.490.3061V,2020ApJ...900...82P}. In the latter case, the fragmentation occurs at a scale much larger than the Jeans length -- a phenomenon that we call super-Jeans fragmentation.  
The key to understanding massive star formation is distinguishing between Jeans and super-Jeans fragmentation. Despite years of research, both Jean-like fragmentation  \citep{2017ApJ...849...25L,2019ApJ...886..102S,2019ApJ...886...36S,2020ApJ...894L..14L} and super-Jeans fragmentation \citep{2009ApJ...696..268Z,2014MNRAS.439.3275W,2018A&A...616L..10F,2023MNRAS.520.3259X}  has been reported, and no conclusion has been reached. 

One key step to distinguish between these scenarios is to infer the gas densities before fragmentation occurs. However, due to the highly complex density distribution of star-forming regions, estimating this initial density distribution is difficult. In previous works, it is assumed that fragmentation takes place in a region of constant density \citep{2020ApJ...894L..14L}, where the fragmentation length does appear to be similar to the Jeans length. When the density is estimated at a scale comparable to the core separation, super-Jeans fragmentation is observed \citep{2023MNRAS.520.3259X}. The key to distinguishing between different fragmentation mechanisms is to construct the initial density distribution.

We combine Delaunay triangulation \citep{Delaunay_1934aa} and Voronoi diagram \citep{Voronoi+1908+97+102} to infer the initial density structure. Delaunay triangulation is an algorithm to construct triangulation meshes from sets of points, and Voronoi diagrams are methods to divide a region based on sets of points. We use Delaunay triangulation to infer the fragmentation length \citep{2021ApJ...916...13L} and combine it with the Voronoi diagram to infer the gas density. This approach allows us to compare the spatial distribution of cores with the predictions of the Jeans criterion with much improved accuracy and test the dependence of fragmentation length on gas density.

\section{Method \& Results}
We use ALMA observations toward three massive star-forming regions in the center of the Galaxy \citep{2020ApJ...894L..14L}. The data have a resolution of $0.25'' \times 0.17''$ (equivalent to $2000 \, \rm AU \times 1400 \, \rm AU$ at a distance of $8.178 \, \rm kpc$ \citealt{2019A&A...625L..10G}), and an image rms measured in emission-free regions without primary-beam corrections of $40 \, \mu \rm Jy \, beam^{-1}$. The sample is 90\% complete to $5 \, M_{\odot}$, and due to an effect called the missing flux, which results from the limited UV coverages of interferometer observations, the mass in the map can be underestimated. These effects can potentially affect our scaling relations. However, as our analyses unfold, one should see that the phenomenon of super-Jeans occurs at the high-mass end, where the cores are complete, and since $l_{\rm Jeans} \propto \rho^{-1/2}$, the conclusion where $l > l_{\rm Jeans}$ should still hold had the missing flux been added in. We use the catalog and the dust continuum map to trace the gas and study fragmentation.

\begin{figure*}
    \includegraphics[width=1 \textwidth]{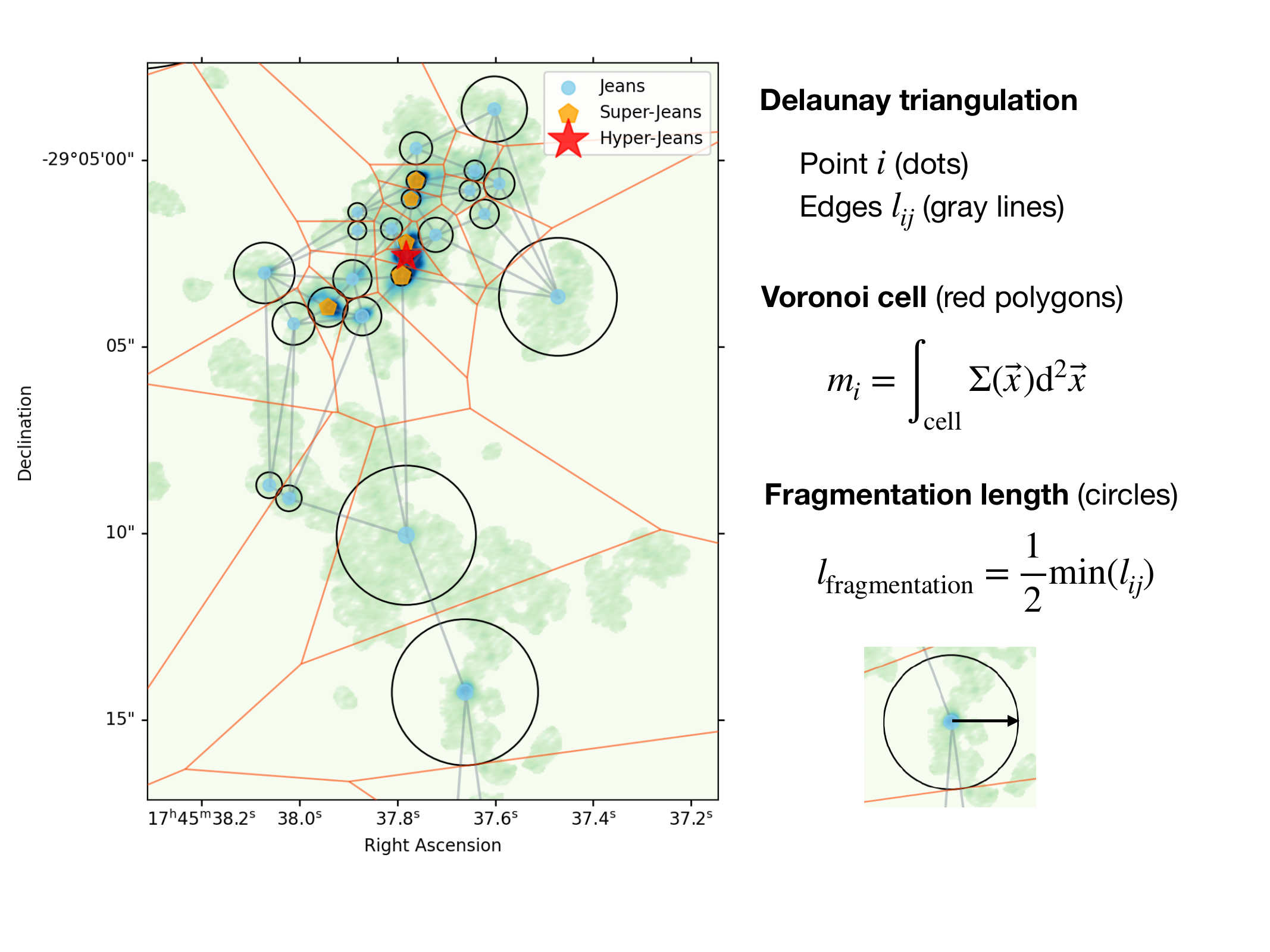}
     \caption{ 
        {\bf Measuring the fragmentation length and mass.} Starting with a catalog
 containing the positions of the cores, we construct the Delaunay
 triangulation (gray lines) and Voronoi diagram (red lines). We use the
 Voronoi diagram to compute the masses of the regions $m_{i, \rm map}$, where $m_i = \int_{\rm cell} \Sigma(\vec{x}) {\rm d} \vec{x}$ 
 and use the Delaunay triangulation to estimate the sizes of the regions $l_i$, where the fragmentation length is one-half of the minimum length of the edges of the Delaunay triangulation starting from the node ($l_{\rm frag} = 1/2 {\rm min} (l_{ij})$)
 (indicated using circles). 
    \label{fig:illus} }
\end{figure*}

\begin{figure*}
    \includegraphics[width=1\textwidth]{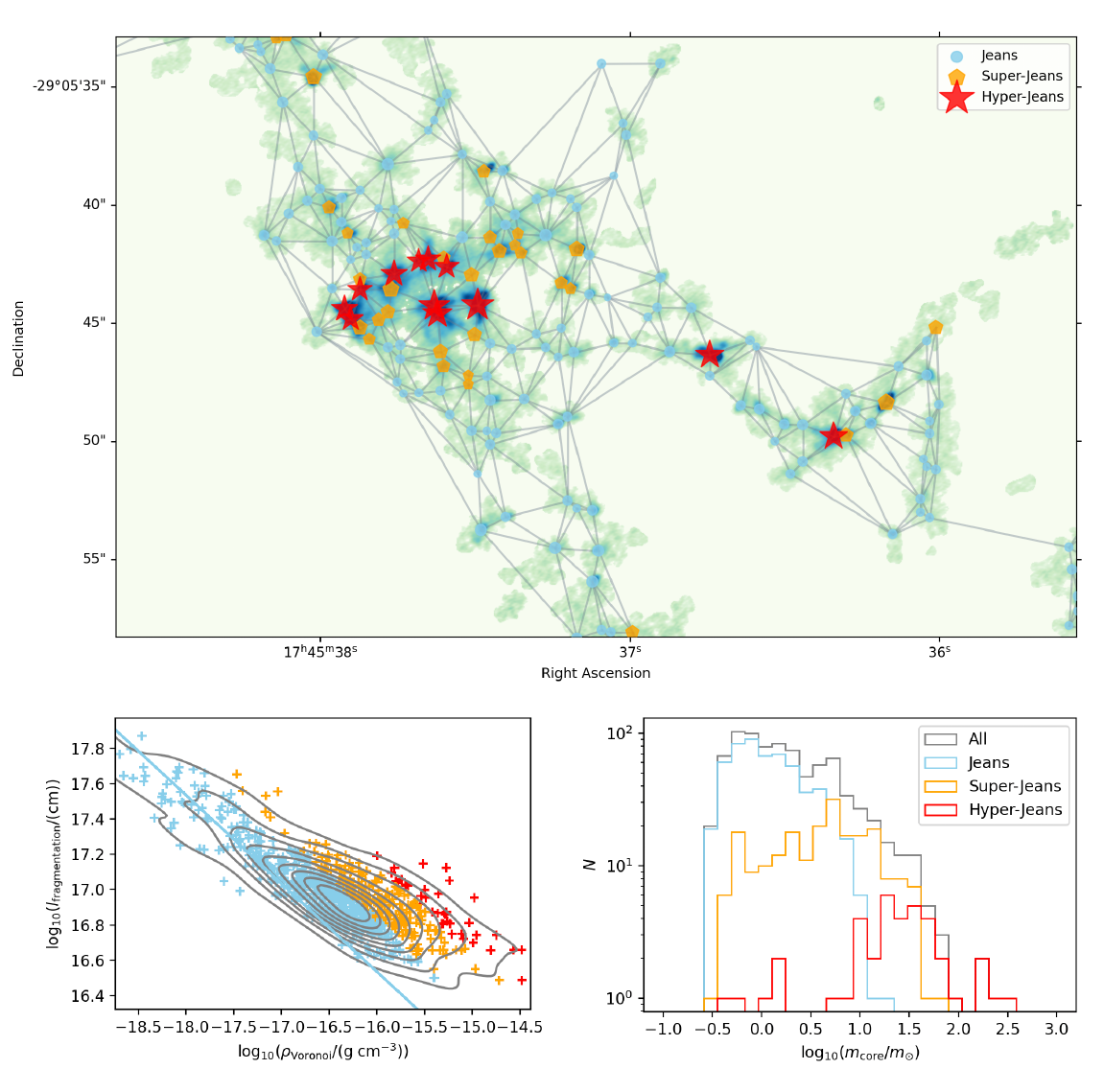}
    \caption{{\bf Jeans, super-Jeans and hyper-Jeans fragmentation}. {\bf Upper panel:} A cutout of the surface density map and spatial distribution of cores in a CMZ cloud called the $20\;\rm km\,s^{-1}$ cloud. The background image is a map of the surface density distribution. Different symbols represent different cores. The lines represent the  Delaunay triangulation, through which the fragmentation length is inferred. Different symbols represent cores produced by Jeans, super-Jeans, and hyper-Jeans fragmentation. {\bf Lower Left panel:} Jeans length plotted against the density. The prediction from the Jeans fragmentation is indicated using the blue solid line. Blue/Yellow/Red symbols represent cores produced through Jeans/Super-Jeans/Hyper-Jeans fragmentation. {\bf Lower Right panel:} Masses spectrum of cores, from the who sample, and those produced through Jeans/Super-Jeans/Hyper-Jeans fragmentation. Results towards other clouds can be found in Appendix \ref{sec:app:d}
    \label{fig:1} }  
\end{figure*}

\begin{table}
    \caption{Summary of different fragmentation modes. $f_{\rm Jeans}$ is defined in Eq. \ref{eq:fjeans}.}\label{tab1}%
    \begin{tabular}{@{}lll@{}}
    \toprule
    Mode & Definition & $M_{\rm core}$ \\
    Jeans fragmentation   & $f_{\rm Jeans} <2$   & $1 M_\odot$ \\
    Super-Jeans fragmentation    & $2<f_{\rm Jeans} <4$   &   $10 M_\odot$ \\
    Hyper-Jeans fragmentation    & $f_{\rm Jeans} <4$    &    $30 M_\odot$ \\
    \bottomrule
    \end{tabular}
    \end{table}

    \subsection{ Estimating  density field  and fragmentation length  }
    To compute the fragmentation scale, we use Delaunay triangulation \citep{Delaunay_1934aa}, a technique for creating a mesh of contiguous, non-overlapping triangles from a dataset of points. The Delaunay triangulation can construct triangular meshes based on the input of a set of points. In our case, similar to \cite{2021ApJ...916...13L}, we use the locations of cores as input, through which the mesh is constructed.

    Towards each data point, the triangulation mesh provides the locations of its closest neighbors. Toward a point (vertex) $p_i$, the triangulation provides the indices and locations of its neighbors. Assuming that the distances to the neighbors are $d_{ij}$, the radius of a region is
    \begin{equation}
     r_{i,\rm Del } = \min_j (d_{ij}) \;,
    \end{equation}
    where $i$ is the point of interest, and $j$ represents its neighbors. The fragmentation length is
    \begin{equation}
     l_{\rm frag} = 2\, r_{\rm Del}\;.
    \end{equation}
    
    To compute the mass, we use the Voronoi diagram \citep{Voronoi+1908+97+102}, which is a method to divide a plane into regions based on a given set of points. In our study, the input points are the locations of the cores.
    For each vertex point $p_i$, the Voronoi diagram provides a region $S_{i}$ which is associated with the region, and the mass of the vertex point is
    \begin{equation}
     m_{i, \rm Vor}= \int_{S_i} \Sigma(x, y) {\rm d} x {\rm d} y
    \end{equation}
    where $\Sigma$ is the surface density distribution. The calculations are performed with \textsc{scipy.spatial} package.
    The definitions of these quantities are illustrated in Fig. \ref{fig:illus}. We note that computation of the $m_{i, \rm Vor}$ is necessary because all the masses contribute to gravity.
    A comparison between the mass of the core $m_{\rm core}$ derived using \textsc{Dendrogram} \citep{2008ApJ...679.1338R}, which is a core extraction algorithm, and the mass of the Voronoi regions $m_{\rm Vor}$ is presented in Fig. \ref{fig:m1m2}. The difference between $m_{\rm core}$ and $m_{\rm Vor}$ is significant, particularly towards low-mass cores. Details on mass calculation, including the conversion from flux to surface density and a step where we remove noisy parts of the map from our analysis, are presented in Appendix \ref{sec:app}.

The initial density can be estimated as \label{eq:rho} 
\begin{equation}
  \rho_{i,\rm  Vor} = \frac{m_{i,\rm  Vor}}{4/3 \pi r_i^3}\;.   
\end{equation}

If the fragmentation is Jeans-like, we expect
\begin{equation}
 l_{i,\rm frag} =  \frac{\pi^{1/2}c_{\rm s}}{\sqrt{G \rho_i}} \;,
\end{equation}
where $c_{\rm s} = \sqrt{k_B T_{\rm gas} / m_{\rm H_2}}$ is the thermal sound speed, $k_{\rm B}$ is the Boltzmann constant, $T_{\rm gas}$ is the temperature of the gas, and $ m_{\rm H_2}$ is the mass of the H$_2$ gas. We assume $T_{\rm gas} = 100$ K, a value agreed upon by the literature \citep{2023ASPC..534...83H}.

In Fig. \ref{fig:1} we plot the relation between density and fragmentation
length. In general, the fragmentation roughly follows the prediction of the Jeans
fragmentation, where a larger density is related to a smaller fragmentation length. In addition, we observe that a significant fraction of
cores with total masses $m_{i, \rm Vor}$ larger than the Jeans mass. To quantify
this excess at high densities, we use the term ``super-Jeans fragmentation" to refer to cores
with $d_{\rm  frag} = 2 \, r_{\rm Del} > 2\, l_{\rm Jeans}$, and ``hyper-Jeans fragmentation" to refer
to cores with $l_{\rm core} > 4\, l_{\rm Jeans}$. Since $m \sim \rho\, l^3$, cores
produced from these fragmentation modes have masses orders of magnitudes larger
than the Jeans mass. These different modes are summarized in Table \ref{tab1}.
From a breakdown of cores under different fragmentation modes (Fig. \ref{fig:1}), where low-mass cores are produced by Jeans fragmentation and cores of higher masses are produced by super and hyper-Jeans fragmentation.



    \begin{figure*}
        \includegraphics[width=1\textwidth]{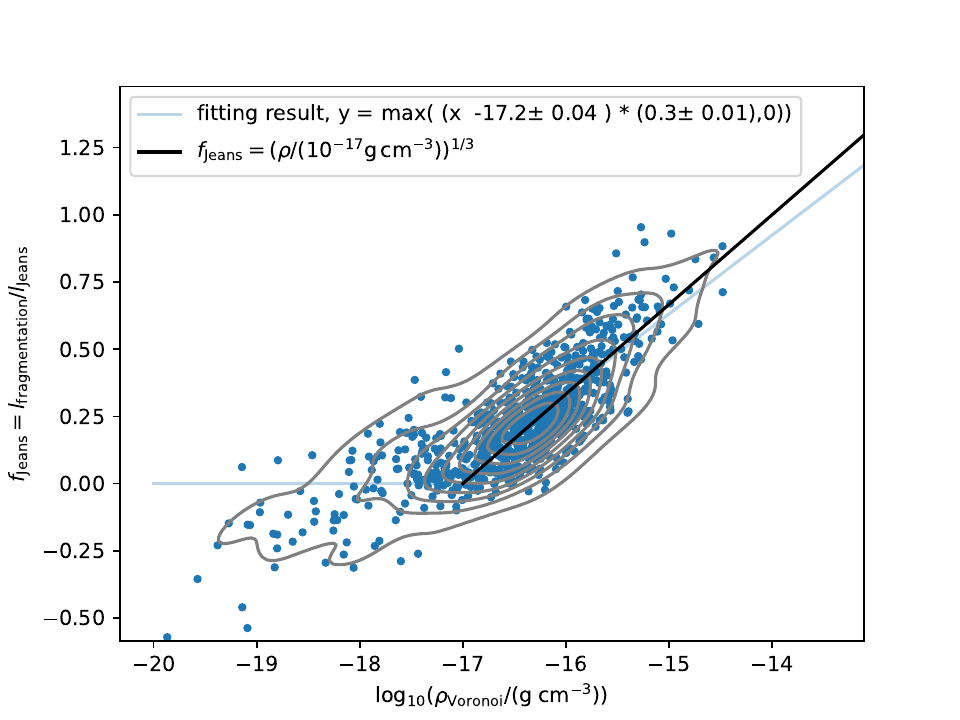}
        \caption{{ \bf Jeans fraction $f_{\rm Jeans} = l_{\rm frag}
        / l_{\rm Jeans}$ as a function of gas density}. The line describes the
        proposed correlation where $f_{\rm Jeans} \propto \rho^{1/3}$. \label{fig:fjeans} }
    \end{figure*}

From a breakdown of core masses in these different categories  (Fig. \ref{fig:1}), 
we find that in the case of CMZ clouds, Jeans-fragmentation dominates the production of $2 \, m_\odot$ cores, super-Jeans fragmentation dominates the production of 10 $m_\odot$ cores, and hyper-Jeans fragmentation dominates the production 30 $m_\odot$ cores. The formation of high-mass cores results from super-Jeans fragmentation.

To quantify the fragmentation, we introduce a parameter called the Jeans ratio, which is the ratio between the observed fragmentation length and the Jeans length, and by plotting 
$f_{\rm Jeans}$ against  $\rho$ (Fig. \ref{fig:fjeans})
we find that the trend can be described by 
{
\begin{equation}\label{eq:fjeans}
 f_{\rm Jeans } = {\rm max}\left( (\frac{\rho}{ 10^{-17}{\rm g} \,{\rm cm}^{-3}})^{1/3},1 \right) \;,
 \end{equation}
}

where the fragmentation mass is
\begin{equation}
 m_{\rm super} =\frac{c_{\rm s}^3}{G^{3/2} \rho^{1/2}}  f_{\rm Jeans}^3 =m_{\rm Jeans} f_{\rm Jeans}^3 \propto \rho^{1/2} \;.
\end{equation}
Since $f_{\rm Jeans}\propto \rho^{1/3}$, we expect $m \propto \rho^{1/2}$, high-mass cores  form at hgh-density environment.


In
Fig. \ref{fig:1}, we use different colors and symbols to represent cores
produced from different fragmentation modes. The super-Jeans fragmentation
tends to occur in groups, with hyper-Jeans cores located at the center and
super-Jeans cores located at the outside. This spatial arrangement suggests a
link between the flow structure at the larger scale and super-Jeans
fragmentation at the center. 


\subsection{Possible theoretical explanations}

There are a few physical mechanisms that can explain the formation of super-Jeans fragmentation. For example, {\color{blue} it is proposed that the fragmentation is delayed,} leading to a more massive object \citep{2019MNRAS.490.3061V,2023arXiv230613846V}, and \citep{2020ApJ...900...82P} proposed an inertia flow mode. Our findings can be tested against simulation results in the future. {\color{blue} \citet{2024MNRAS.528.7333L}  derived non-stationary corrections, where the Jeans ratio becomes }

\begin{equation}
 f_{\rm Jeans} = (t_{\rm ff} / t_{\rm acc})^{1/3} \;,
\end{equation}
where $t_{\rm acc} = \rho / \dot \rho$ under the non-stationary case. A positive $\dot \rho$ leads to a larger fragmentation length. 

Letting 
\begin{equation}
 f_{\rm Jeans }  \propto \rho^{1/3} \;,
\end{equation}
we expect
\begin{equation}
 t_{\rm acc} \propto \rho^{-1.5}, \dot \rho \propto \rho^{2.5} \;,
\end{equation}
where {\color{blue} the high-density region should have short $t_{\rm acc}$} to ensure super-Jeans fragmentation. The super-Jeans fragmentation can be related to the dynamic nature of the high-density regions. This can be explained if the whole region is collapsing globally. Super and hyper-Jeans fragmentation occurs at the central regions where the flow converges, and Jeans fragmentation occurs at the outside. Other processes such as core accretion and further evolution of core separations, can also move the cores around in the $\rho_{\rm Vor}$-$l_{\rm \rm frag}$ diagram (Fig. \ref{fig:1}), and can potentially explain the super-Jeans accretion.

{\subsection{Line-of-sight effects}
We note that the analyses are performed in 2D, whereas the density on which the Jeans length depends is a quantity that can only be measured in 3D. Thus, the analysis requires that the sizes of the regions can be estimated using the separations of the adjacent cores, as well as the fact that the line of sight effects are not significant. These assumptions should hold since the cloud can be approximated as a set of nested regions of significant density contrast, e.g., $\rho \sim r^{-2}$ or $r^{-3}$ \citep{2022MNRAS.514L..16L}. This is discussed in detail in Appendix \ref{sec:c}. One should also note that massive cores have separations that are a few times the Jeans length. For these to be explained by line-of-sight effects, we require the mass to be overestimated by a factor of 10, which is highly unlikely.
}

\section{Conclusion}
The fragmentation of gas clouds and the formation of massive stars is one of the
most complex astrophysical processes. The Jeans criterion sets the
foundation of gravitational fragmentation, yet the observational results remain inconclusive.
Fragmentation is the process that links an initial density distribution, which
is produced by turbulence and global collapse, to a set of regions that collapse on their own. Previous studies assumed flat initial density
distributions, neglecting the internal
density distributions. By developing a new adaptive approach based on the
Delaunay triangulation and Voronoi diagram, we estimate the initial density structure and study gravitational fragmentation with much-improved accuracy.

We find that at the low-density end, the fragmentation is Jeans-like;
at the high-density end, we observe a significant fraction of massive cores
produced by fragmentation whose scales are much larger than the Jeans
fragmentation. We use $f_{\rm Jeans} =
l_{\rm frag} / l_{\rm Jeans}$ to describe deviations from the
Jeans fragmentation and find that $f_{\rm Jeans} \propto \rho ^{1/3}$.


The paper demonstrates the power of numerical methods developed in computational geometry, such as Delaunay triangulation and Voronoi diagram, in analyzing complex patterns that emerged in modern astrophysical research, where effective ways to quantify these complex structures is a key step towards physical understanding.

\section*{Data availability}
Table data: \url{doi: 10.3847/
2041-8213/ab8b65}. Image data: \url{https://doi.org/10.5281/zenodo.4740418}.

\section*{Acknowledgements}
GXL acknowledges support from NSFC grant No. 12273032 and
12033005. X.L.\ acknowledges support from the National Key R\&D Program of China (No.\ 2022YFA1603101), the Strategic Priority Research Program of the Chinese Academy of Sciences (CAS) Grant No.\ XDB0800300, the National Natural Science Foundation of China (NSFC) through grant Nos.\ 12273090 and 12322305, the Natural Science Foundation of Shanghai (No.\ 23ZR1482100), and the CAS ``Light of West China'' Program No.\ xbzg-zdsys-202212.

\bibliography{paper}

\begin{thebibliography}{}
\expandafter\ifx\csname natexlab\endcsname\relax\def\natexlab#1{#1}\fi
\providecommand{\url}[1]{\href{#1}{#1}}
\providecommand{\dodoi}[1]{doi:~\href{http://doi.org/#1}{\nolinkurl{#1}}}
\providecommand{\doeprint}[1]{\href{http://ascl.net/#1}{\nolinkurl{http://ascl.net/#1}}}
\providecommand{\doarXiv}[1]{\href{https://arxiv.org/abs/#1}{\nolinkurl{https://arxiv.org/abs/#1}}}

\bibitem[{{Bonnell} {et~al.}(2001){Bonnell}, {Bate}, {Clarke}, \& {Pringle}}]{2001MNRAS.323..785B}
{Bonnell}, I.~A., {Bate}, M.~R., {Clarke}, C.~J., \& {Pringle}, J.~E. 2001, \mnras, 323, 785, \dodoi{10.1046/j.1365-8711.2001.04270.x}

\bibitem[{Delaunay(1934)}]{Delaunay_1934aa}
Delaunay, B. 1934, Bulletin de l'Acadeemie des Sciences de l'URSS. Classe des sciences mathematiques et na, 1934, 793

\bibitem[{{Figueira} {et~al.}(2018){Figueira}, {Bronfman}, {Zavagno}, {Louvet}, {Lo}, {Finger}, \& {Rod{\'o}n}}]{2018A&A...616L..10F}
{Figueira}, M., {Bronfman}, L., {Zavagno}, A., {et~al.} 2018, \aap, 616, L10, \dodoi{10.1051/0004-6361/201832930}

\bibitem[{{GRAVITY Collaboration} {et~al.}(2019){GRAVITY Collaboration}, {Abuter}, {Amorim}, {Baub{\"o}ck}, {Berger}, {Bonnet}, {Brandner}, {Cl{\'e}net}, {Coud{\'e} Du Foresto}, {de Zeeuw}, {Dexter}, {Duvert}, {Eckart}, {Eisenhauer}, {F{\"o}rster Schreiber}, {Garcia}, {Gao}, {Gendron}, {Genzel}, {Gerhard}, {Gillessen}, {Habibi}, {Haubois}, {Henning}, {Hippler}, {Horrobin}, {Jim{\'e}nez-Rosales}, {Jocou}, {Kervella}, {Lacour}, {Lapeyr{\`e}re}, {Le Bouquin}, {L{\'e}na}, {Ott}, {Paumard}, {Perraut}, {Perrin}, {Pfuhl}, {Rabien}, {Rodriguez Coira}, {Rousset}, {Scheithauer}, {Sternberg}, {Straub}, {Straubmeier}, {Sturm}, {Tacconi}, {Vincent}, {von Fellenberg}, {Waisberg}, {Widmann}, {Wieprecht}, {Wiezorrek}, {Woillez}, \& {Yazici}}]{2019A&A...625L..10G}
{GRAVITY Collaboration}, {Abuter}, R., {Amorim}, A., {et~al.} 2019, \aap, 625, L10, \dodoi{10.1051/0004-6361/201935656}

\bibitem[{{Henshaw} {et~al.}(2023){Henshaw}, {Barnes}, {Battersby}, {Ginsburg}, {Sormani}, \& {Walker}}]{2023ASPC..534...83H}
{Henshaw}, J.~D., {Barnes}, A.~T., {Battersby}, C., {et~al.} 2023, in Astronomical Society of the Pacific Conference Series, Vol. 534, Protostars and Planets VII, ed. S.~{Inutsuka}, Y.~{Aikawa}, T.~{Muto}, K.~{Tomida}, \& M.~{Tamura}, 83, \dodoi{10.48550/arXiv.2203.11223}

\bibitem[{{Jeans}(1902)}]{1902RSPTA.199....1J}
{Jeans}, J.~H. 1902, Philosophical Transactions of the Royal Society of London Series A, 199, 1, \dodoi{10.1098/rsta.1902.0012}

\bibitem[{{Kainulainen} {et~al.}(2014){Kainulainen}, {Federrath}, \& {Henning}}]{2014Sci...344..183K}
{Kainulainen}, J., {Federrath}, C., \& {Henning}, T. 2014, Science, 344, 183, \dodoi{10.1126/science.1248724}

\bibitem[{{Li}(2022)}]{2022ApJS..259...59L}
{Li}, G.-X. 2022, \apjs, 259, 59, \dodoi{10.3847/1538-4365/ac4bc4}

\bibitem[{{Li}(2024{\natexlab{a}})}]{2024MNRAS.528.7333L}
---. 2024{\natexlab{a}}, \mnras, 528, 7333, \dodoi{10.1093/mnras/stae384}

\bibitem[{{Li}(2024{\natexlab{b}})}]{2024MNRAS.528L..52L}
---. 2024{\natexlab{b}}, \mnras, 528, L52, \dodoi{10.1093/mnrasl/slad149}

\bibitem[{{Li} \& {Burkert}(2017)}]{2017MNRAS.464.4096L}
{Li}, G.-X., \& {Burkert}, A. 2017, \mnras, 464, 4096, \dodoi{10.1093/mnras/stw2504}

\bibitem[{{Li} {et~al.}(2021){Li}, {Cao}, \& {Qiu}}]{2021ApJ...916...13L}
{Li}, G.-X., {Cao}, Y., \& {Qiu}, K. 2021, \apj, 916, 13, \dodoi{10.3847/1538-4357/ac01d4}

\bibitem[{{Li} \& {Zhou}(2022)}]{2022MNRAS.514L..16L}
{Li}, G.-X., \& {Zhou}, J.-X. 2022, \mnras, 514, L16, \dodoi{10.1093/mnrasl/slac049}

\bibitem[{{Liu} {et~al.}(2017){Liu}, {Lacy}, {Li}, {Wang}, {Qin}, {Zhang}, {Kim}, {Garay}, {Wu}, {Mardones}, {Zhu}, {Tatematsu}, {Hirota}, {Ren}, {Liu}, {Chen}, {Su}, \& {Li}}]{2017ApJ...849...25L}
{Liu}, T., {Lacy}, J., {Li}, P.~S., {et~al.} 2017, \apj, 849, 25, \dodoi{10.3847/1538-4357/aa8d73}

\bibitem[{{Lu} {et~al.}(2020){Lu}, {Cheng}, {Ginsburg}, {Longmore}, {Kruijssen}, {Battersby}, {Zhang}, \& {Walker}}]{2020ApJ...894L..14L}
{Lu}, X., {Cheng}, Y., {Ginsburg}, A., {et~al.} 2020, \apjl, 894, L14, \dodoi{10.3847/2041-8213/ab8b65}

\bibitem[{{Ossenkopf} \& {Henning}(1994)}]{1994A&A...291..943O}
{Ossenkopf}, V., \& {Henning}, T. 1994, \aap, 291, 943

\bibitem[{{Padoan} {et~al.}(2020){Padoan}, {Pan}, {Juvela}, {Haugb{\o}lle}, \& {Nordlund}}]{2020ApJ...900...82P}
{Padoan}, P., {Pan}, L., {Juvela}, M., {Haugb{\o}lle}, T., \& {Nordlund}, {\r{A}}. 2020, \apj, 900, 82, \dodoi{10.3847/1538-4357/abaa47}

\bibitem[{{Rosolowsky} {et~al.}(2008){Rosolowsky}, {Pineda}, {Kauffmann}, \& {Goodman}}]{2008ApJ...679.1338R}
{Rosolowsky}, E.~W., {Pineda}, J.~E., {Kauffmann}, J., \& {Goodman}, A.~A. 2008, \apj, 679, 1338, \dodoi{10.1086/587685}

\bibitem[{{Sanhueza} {et~al.}(2019){Sanhueza}, {Contreras}, {Wu}, {Jackson}, {Guzm{\'a}n}, {Zhang}, {Li}, {Lu}, {Silva}, {Izumi}, {Liu}, {Miura}, {Tatematsu}, {Sakai}, {Beuther}, {Garay}, {Ohashi}, {Saito}, {Nakamura}, {Saigo}, {Veena}, {Nguyen-Luong}, \& {Tafoya}}]{2019ApJ...886..102S}
{Sanhueza}, P., {Contreras}, Y., {Wu}, B., {et~al.} 2019, \apj, 886, 102, \dodoi{10.3847/1538-4357/ab45e9}

\bibitem[{{Svoboda} {et~al.}(2019){Svoboda}, {Shirley}, {Traficante}, {Battersby}, {Fuller}, {Zhang}, {Beuther}, {Peretto}, {Brogan}, \& {Hunter}}]{2019ApJ...886...36S}
{Svoboda}, B.~E., {Shirley}, Y.~L., {Traficante}, A., {et~al.} 2019, \apj, 886, 36, \dodoi{10.3847/1538-4357/ab40ca}

\bibitem[{{V{\'a}zquez-Semadeni} {et~al.}(2023){V{\'a}zquez-Semadeni}, {G{\'o}mez}, \& {Gonz{\'a}lez-Samaniego}}]{2023arXiv230613846V}
{V{\'a}zquez-Semadeni}, E., {G{\'o}mez}, G.~C., \& {Gonz{\'a}lez-Samaniego}, A. 2023, arXiv e-prints, arXiv:2306.13846, \dodoi{10.48550/arXiv.2306.13846}

\bibitem[{{V{\'a}zquez-Semadeni} {et~al.}(2019){V{\'a}zquez-Semadeni}, {Palau}, {Ballesteros-Paredes}, {G{\'o}mez}, \& {Zamora-Avil{\'e}s}}]{2019MNRAS.490.3061V}
{V{\'a}zquez-Semadeni}, E., {Palau}, A., {Ballesteros-Paredes}, J., {G{\'o}mez}, G.~C., \& {Zamora-Avil{\'e}s}, M. 2019, \mnras, 490, 3061, \dodoi{10.1093/mnras/stz2736}

\bibitem[{Voronoi(1908)}]{Voronoi+1908+97+102}
Voronoi, G. 1908, Journal für die reine und angewandte Mathematik (Crelles Journal), 1908, 97, \dodoi{doi:10.1515/crll.1908.133.97}

\bibitem[{{Wang} {et~al.}(2014){Wang}, {Zhang}, {Testi}, {van der Tak}, {Wu}, {Zhang}, {Pillai}, {Wyrowski}, {Carey}, {Ragan}, \& {Henning}}]{2014MNRAS.439.3275W}
{Wang}, K., {Zhang}, Q., {Testi}, L., {et~al.} 2014, \mnras, 439, 3275, \dodoi{10.1093/mnras/stu127}

\bibitem[{{Xu} {et~al.}(2023){Xu}, {Wang}, {Liu}, {Goldsmith}, {Zhang}, {Juvela}, {Liu}, {Qin}, {Li}, {Tej}, {Garay}, {Bronfman}, {Li}, {Wu}, {G{\'o}mez}, {V{\'a}zquez-Semadeni}, {Tatematsu}, {Ren}, {Zhang}, {Toth}, {Liu}, {Yue}, {Zhang}, {Baug}, {Issac}, {Stutz}, {Liu}, {Fuller}, {Tang}, {Zhang}, {Dewangan}, {Lee}, {Zhou}, {Xie}, {Jiao}, {Wang}, {Liu}, {Luo}, {Soam}, \& {Eswaraiah}}]{2023MNRAS.520.3259X}
{Xu}, F.-W., {Wang}, K., {Liu}, T., {et~al.} 2023, \mnras, 520, 3259, \dodoi{10.1093/mnras/stad012}

\bibitem[{{Zhang} {et~al.}(2009){Zhang}, {Wang}, {Pillai}, \& {Rathborne}}]{2009ApJ...696..268Z}
{Zhang}, Q., {Wang}, Y., {Pillai}, T., \& {Rathborne}, J. 2009, \apj, 696, 268, \dodoi{10.1088/0004-637X/696/1/268}

\end{thebibliography}

\bibliographystyle{aasjournal}



\appendix

\section{Gas mass measurement}\label{sec:app}
We use the dust continuum map from an ALMA observation towards the center of our Galaxy \citep{2020ApJ...894L..14L} to derive the total mass of the fragments. We compute the surface density map, we use the equation
\begin{equation}
   \Sigma = R \frac{F_{\nu}}{B_{\nu}(T) \kappa_{\nu}} \;,
\end{equation}
where $F_{\nu}$ is the flx density, $B_{\nu}(T)$ is the Planck function  and $\kappa$ is the dust emissivity, and  $R$ is the gas-to-dust mass ratio.  We assume $R=100$, $\kappa_\nu =\rm 0.9\, g\;\rm cm^{-1}$ \citep{1994A&A...291..943O}.

\begin{figure*}
    \includegraphics[width=1\textwidth]{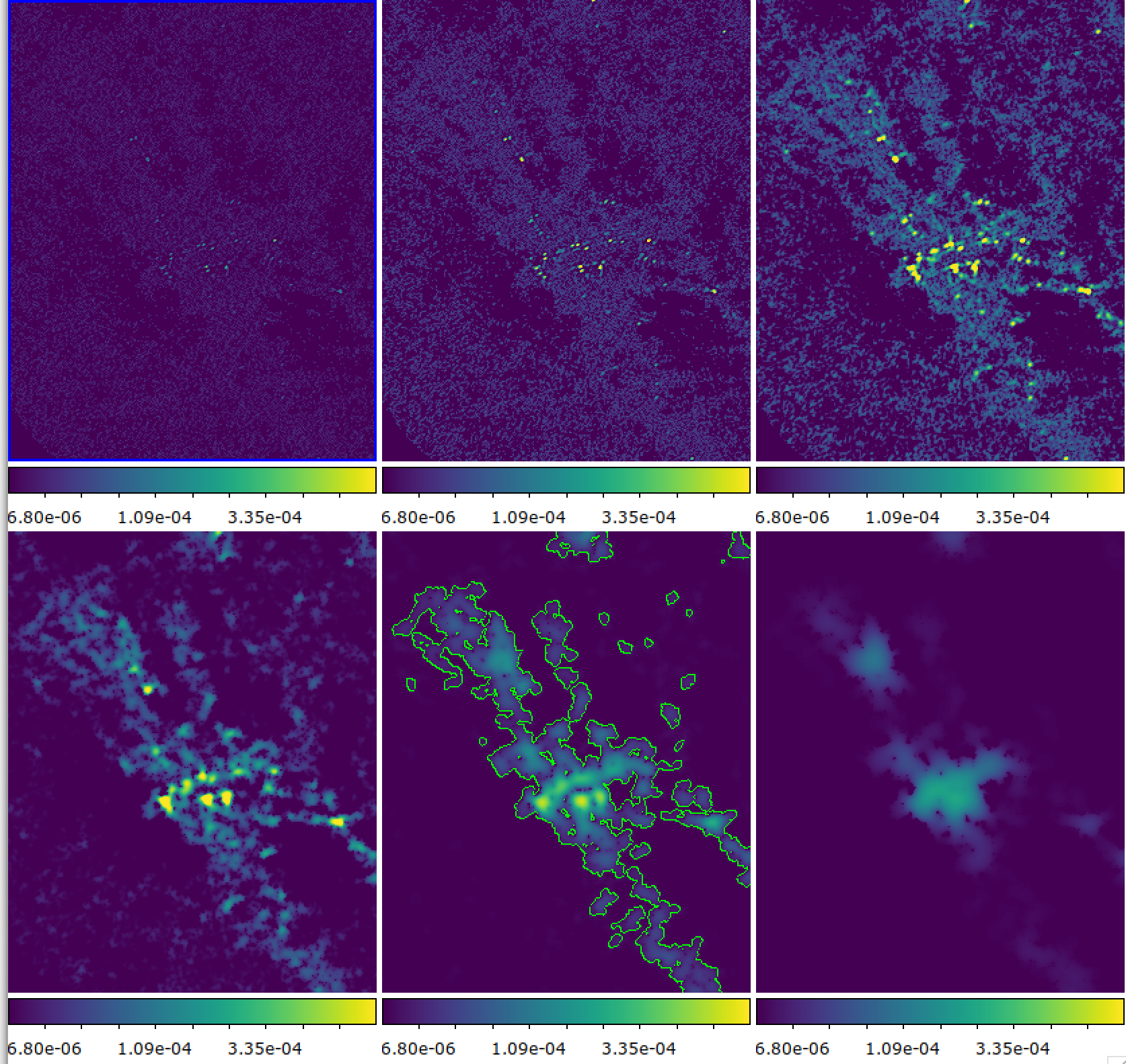}
    \caption{\label{fig:diffuse} { Method of creating a mask where most of the significant emission is include}. We first decompose the map using the method of \textsc{constrained diffusion decomposition} \citep{2022ApJS..259...59L}. Then the a is created at the 4th component, at the threshold of 1 mJy\,Beam$^{-1}$. The boundaries of this mask is indicated using green contours in the 5th plot. The data shown here is taken at observations towards 20 km/s cloud from \citep{2020ApJ...894L..14L}. }
\end{figure*}

Note that the map contains a significant amount of correlated noise in regions
of low signal-to-noise ratios. To achieve improved accuracy, we remove these empty regions from our analysis by creating a mask where most of the emission is included. To achieve this, we use the method of
\textsc{constrained diffusion decomposition} \citep{2022ApJS..259...59L} to decompose the
input maps into component maps that contain structures at different scales, and
the threshold is selected at the $n=4$ component (which has a resolution of $2^4 = 16$ pixels) at a threshold of 1, 2, 1, 2
$\rm mJy\,Beam^{-1}$ for observations towards the 20 km/s cloud, 50 km/s cloud, SgrC, and SgrB1 respectively.
This procedure is illustrated in Fig \ref{fig:diffuse}.

\section{Mass corrections}
\begin{figure*}
    \includegraphics[width=1\textwidth]{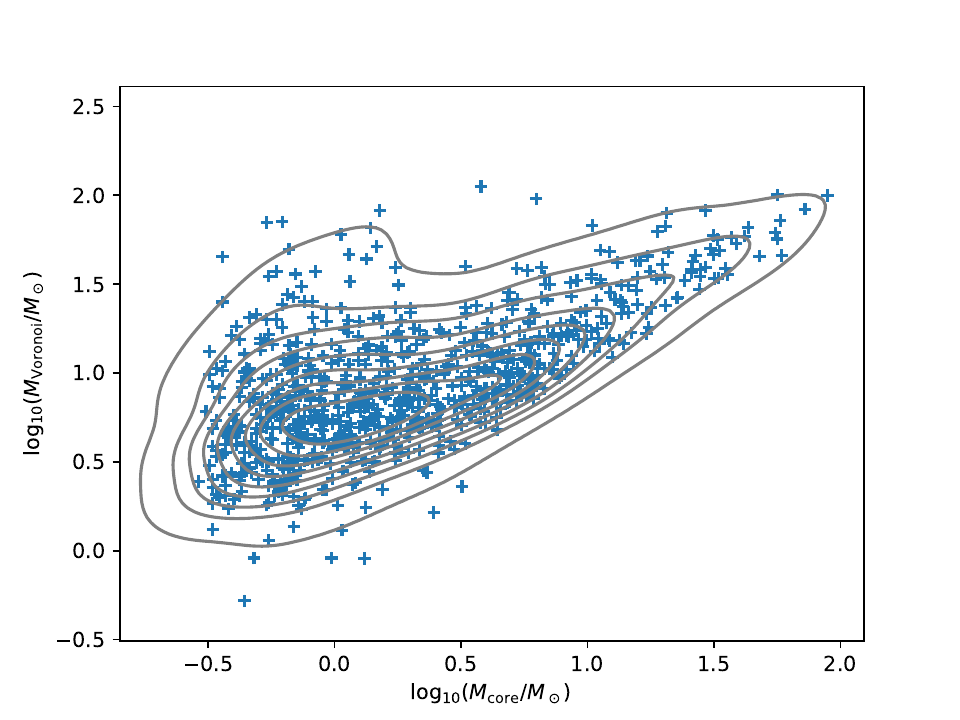}
    \caption{ 
    { Fragmentation mass $m_{\rm map}$ plotted against the core mass $m_{\rm core}$.}    
    Fragmentation mass $m_{\rm map}$ derived by summing over all the region partition produced using the Voronoi diagram, plotted against the masses of cores $m_{\rm core}$. The differences between these two are caused by diffuse emissions outside the cores. \label{fig:m1m2} }
\end{figure*}

In Fig. \ref{fig:m1m2} we present a comparison between the mass of the core $m_{\rm core}$, as derived in Lu et al. 2020 \citep{2020ApJ...894L..14L},  with the mass derived by summing over all the emission contained in the corresponding region in the Voronoi diagram $v_{\rm Vor}$. The mass in the Voronoi diagram region is significant, particularly towards cores of smaller masses.  

\section{Line-of-sight effect on density estimation}\label{sec:c}
A critical assumption in our analysis is that the gas density can be estimated as (Eq. \ref{eq:rho})
\begin{equation}
    \rho_{i,\rm Vor} = \frac{m_{i,\rm Vor}}{4/3 \pi r_i^3} = \frac{\Sigma_{i,\rm Vor}}{4/3 \pi r_i}  \;.
\end{equation}
where $\Sigma_{i,\rm Vor} = m_{i,\rm Vor} r_i^{-2}$. For the density estimates to hold, the volume density is expected to be related to the surface density via
\begin{eqnarray} \label{eq:rho:approx}
    \rho \approx \frac{\Sigma_{i,\rm Vor}}{r_i}\;.
\end{eqnarray}
Although we do not expect Eq. \ref{eq:rho:approx} to hold in all possible configurations, it is likely to be a good approximation when the cloud can be considered as a set of density enhancements nested within one another, with high-density structures staying at the center.

{\color{blue} The key to our density estimation is that the clouds we study should have centrally-condensed density structures. Consider a cloud which has a power-law density PDF: $ P(\Sigma) \propto \Sigma^{-2}$. The surface area occupied by the high-density region is much smaller than the area occupied by the low-density region $(\Sigma \propto \Sigma^{-1})$. This has two consequences: First, a dense gas has a low chance to overlap spatially, unless physically associated, and second, the contributions from the low-density envelope are not significant. This fact, where one can estimate the density by combining surface density measurements with thickness estimations, has been used in a number of papers \citep{2014Sci...344..183K,2017MNRAS.464.4096L,2022MNRAS.514L..16L,2024MNRAS.528L..52L}.}

\begin{figure}
    \includegraphics[width = 0.5 \textwidth]{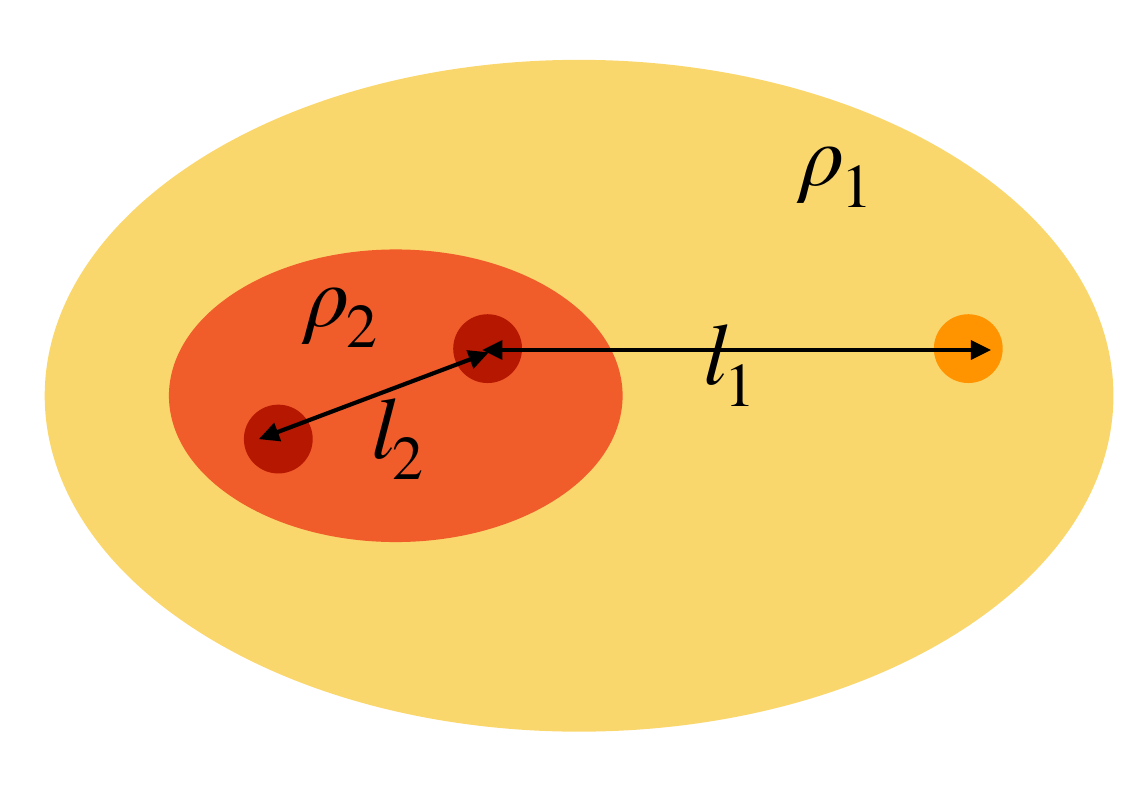}
    \caption{{\bf A illustration of the "cloud-in-cloud" scenario.}, We consider the fragmentation of the cloud, with a high-density region, or density $\rho_2$ and size $l_2$, embedded in a region larger region of density $\rho_1$ and size $l_1$. The effect of the presence of a low-density region, 1, on the estimation of the density of region 2, is minimal. \label{fig:cloud}}
\end{figure}

To further illustrate this, we consider a density configuration as in Fig. \ref{fig:cloud}, where we consider the fragmentation of a cloud, which consists of a high-density region, of density $\rho_2$, embedded in an envelope of lower density $\rho_1$. Region 2 has fragmented into some cores, with a typical separation of $l_2$. Our density estimation holds as long as the separation $l_2$ is comparable to the size of the region, as well as there being some significant density contrast between the high- and low-density regions. Molecular clouds are complex objects whose density PDF can be modeled using, e.g., log-normal or power-law distributions, which contain significant density fluctuations on all observable scales. By decomposing the cloud using algorithms such as \textsc{Dendrogram}, one often observes that one coherent region often contains only a small number of fragments, such that the separation between the fragments is comparable to the size of the region. Another assumption is that the foreground and background do not affect the density estimation. In this case, towards region 2, the contribution of the gas from region 1 is
\begin{equation}
    \rho_{\rm contamination} \approx \frac{\rho_1 l_1}{l_2}\;.
\end{equation}
Assuming a power-law relation between region size and density, e.g.,
\begin{equation}
 \frac{\rho_1}{\rho_2} \approx \left( \frac{l_1}{l_2} \right)^{-2}\;,
\end{equation}
the fractional contribution from the line-of-sight effect is
\begin{equation}
    \delta_\rho = \frac{\rho_{\rm contamination}}{\rho_2} = \frac{l_2}{l_1} \ll 1   \;,
\end{equation}
when $l_2 \ll l_1$. The line-of-sight effects are negligible as long as there are some significant size separations.

{\color{blue}
Of course, we must acknowledge that if the density structure is singular, e.g., if we view a filament from its side, estimating the density by combining surface density and scale information can lead to incorrect results. However, seeing a straight filament from its side is extremely unlikely, as filaments have moderate aspect ratios and are often bent. Even if we have a filament that is fully straight, the chance of observing it from the side is quite low. We thus conclude that our density estimate should be accurate, although not without caveats.
}

\section{Plots for all regions}\label{sec:app:d}
\begin{figure*}
    \includegraphics[width=1\textwidth]{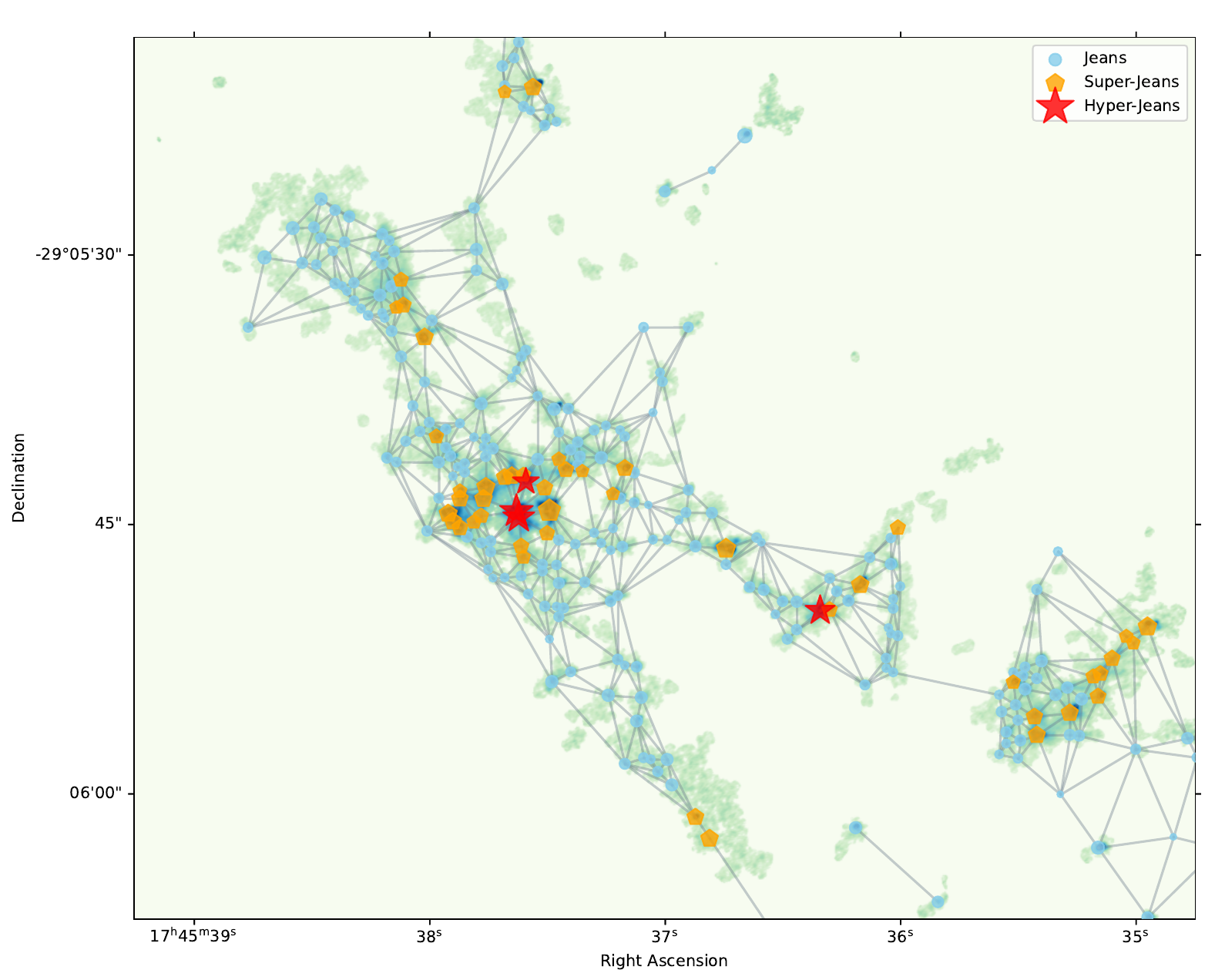}
    \caption{{\bf Surface density map and fragmentation of the 20 $\;\rm km \, s^{-1}$ cloud.} The background image is a map of the surface density distribution. Different symbols represent different cores. The lines represent the  Delaunay triangulation, through which the fragmentation length is inferred. Different symbols represent cores produced by Jeans, super-Jeans, and hyper-Jeans fragmentation. 
    \label{fig:a1} }  
\end{figure*}

\begin{figure*}
    \includegraphics[width=1\textwidth]{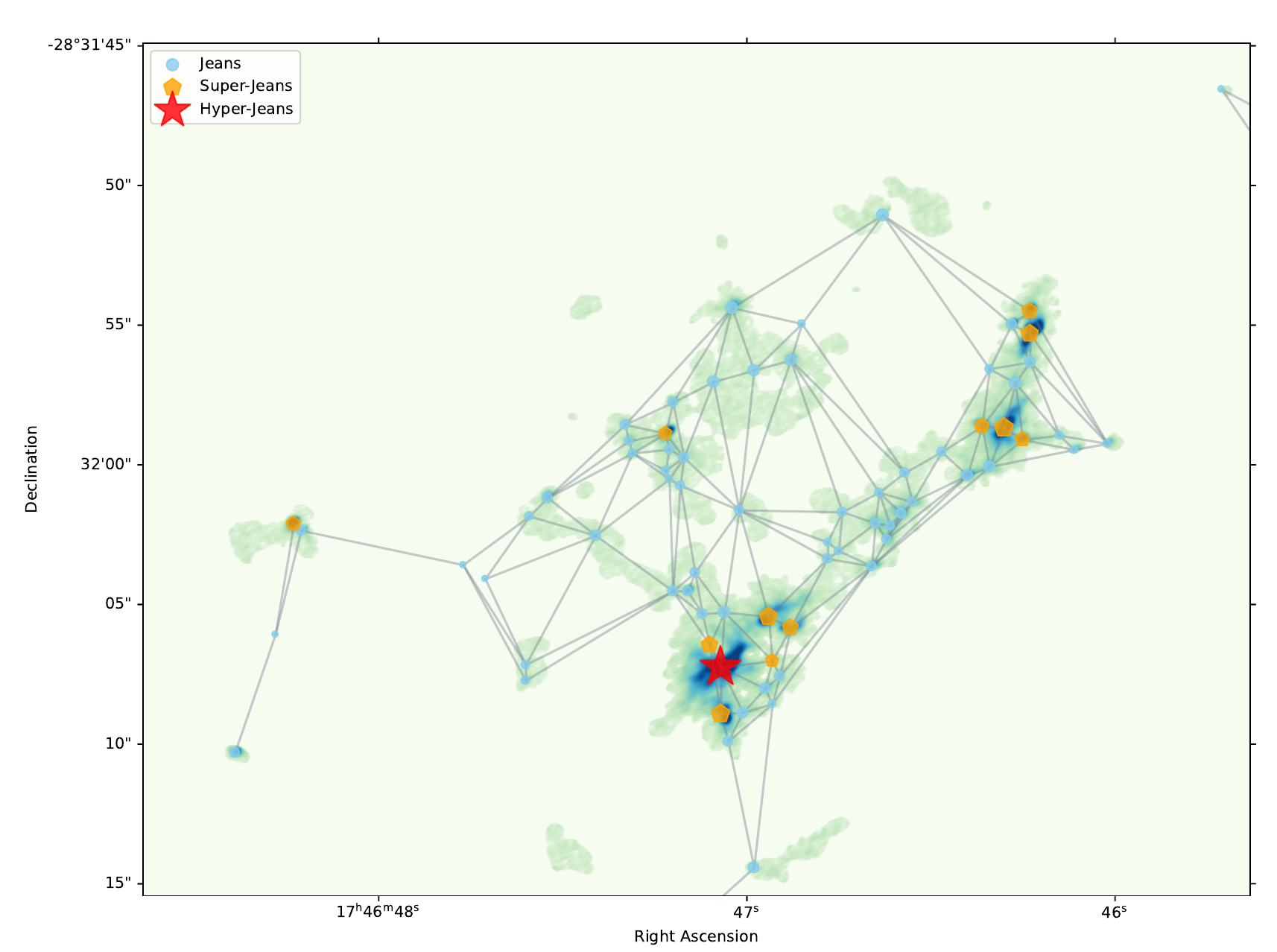}
    \caption{{\bf Surface density map and fragmentation of the Sgr B1 off cloud.}. The background image is a map of the surface density distribution. Different symbols represent different cores. The lines represent the  Delaunay triangulation, through which the fragmentation length is inferred. Different symbols represent cores produced by Jeans, super-Jeans, and hyper-Jeans fragmentation. 
    \label{fig:a2} }  
\end{figure*}

\begin{figure*}
    \includegraphics[width=1\textwidth]{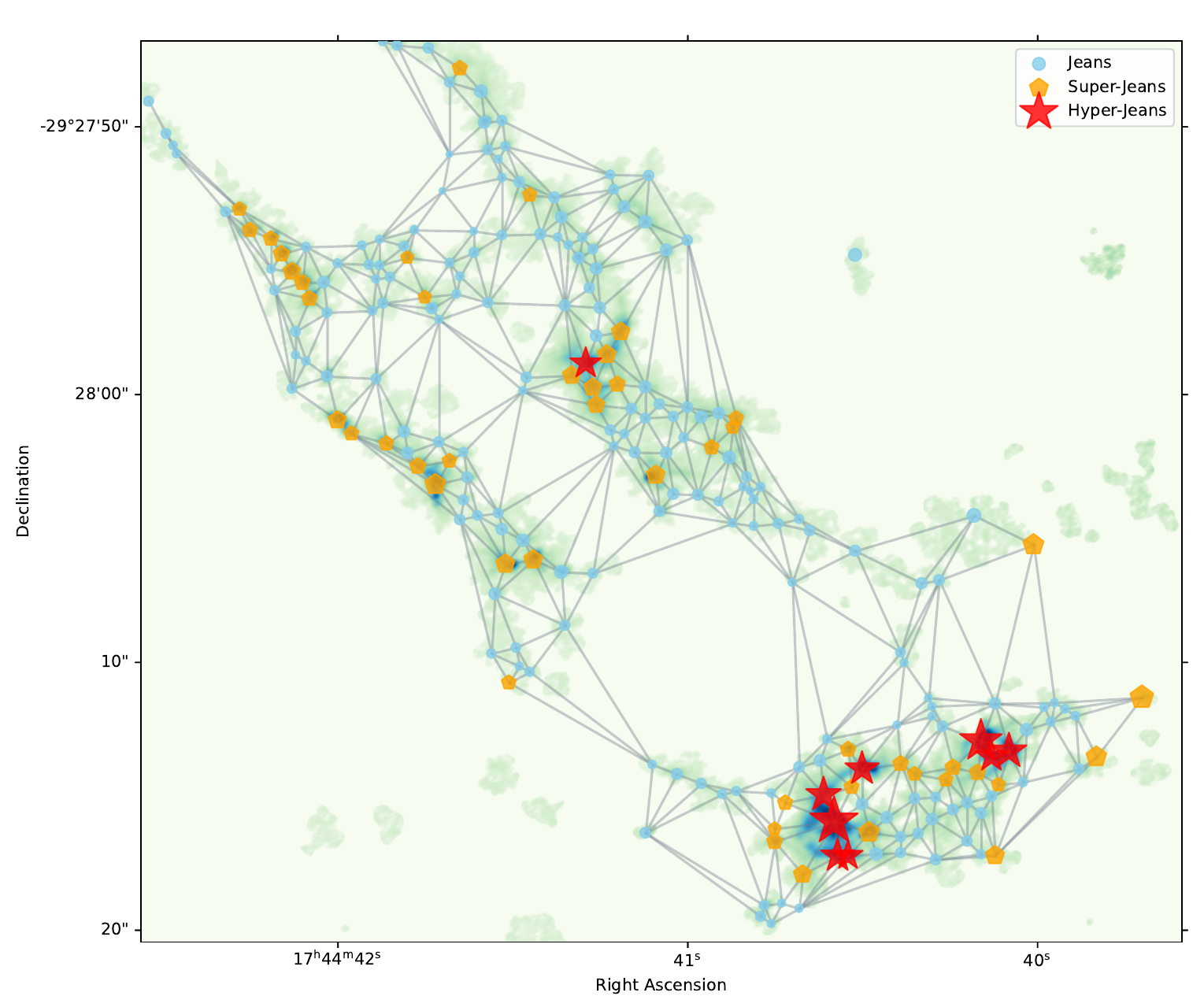}
    \caption{{\bf Surface density map and fragmentation of the Sgr C cloud.}. The background image is a map of the surface density distribution. Different symbols represent different cores. The lines represent the  Delaunay triangulation, through which the fragmentation length is inferred. Different symbols represent cores produced by Jeans, super-Jeans, and hyper-Jeans fragmentation. 
    \label{fig:a3} }  
\end{figure*}

In Figs \ref{fig:a1}, \ref{fig:a2} and \ref{fig:a3}, we present the surface density maps and fragmentation of the three clouds. 

\bsp	
\label{lastpage}
\end{document}